# Effect of lamellar nanostructures on the second harmonic generation of polymethylmethacrylate films doped with 4-(4-nitrophenylazo)aniline chromophores

Alfredo Franco[a,b*], Laura Romero-Miranda[a], Guadalupe Valverde-Aguilar[a], Jorge A. García-Macedo[a], Giovanna Brusatin[b], Massimo Guglielmi[b]

[a]Departamento de Estado Sólido, Instituto de Física, Universidad Nacional Autónoma de México. 04510 México, D.F.
[b]Dipartimento di Ingegneria Meccanica, Settore Materiali, Università di Padova, via Marzolo 9, 35131 Padova, Italy.

## ABSTRACT

The kinetics of the orientation of Disperse Orange 3 molecules embedded in amorphous and nanostructured Polymethylmethacrylate films was studied under the effect of an intense electrostatic poling field. Non-centrosymmetric chromophore distributions were obtained in Polymethylmethacrylate films by Corona poling technique. These distributions depends on the Corona poling time. The changes in the orientation of the Disperse Orange 3 molecules were followed by in-situ transmitted Second Harmonic Generation measurements. The Second Harmonic Generation signal was recorded as function of time at several temperatures; it was fitted as function of the Corona poling time, considering matrix-chromophore interactions. The Polymethylmethacrylate films were nanostructured by the incorporation of an anionic surfactant, the Sodium Dodecyl Sulfate. The lamellar nanostructures in the films were identified by X-ray diffraction measurements.

**HIGHLIGHTS:** Study of the kinetics of orientation of Disperse Orange 3 chromophores in nanostructured Polymethylmetacrylate films. The nanostructures were identified by X-ray diffraction measurements. Non-centrosymmetric chromophore orientation distributions were obtained by Corona poling technique. Chromophores orientation was followed by Second Harmonic Generation as function of poling time at several temperatures. In the nanostructured films the orientation of the chromophores is faster and their disorientation is slower.

## 1. INTRODUCTION

The development of polymers as materials for non-linear optics (NLO) has largely extended the range of applications of organic materials in optoelectronics. Composite polymeric materials are indeed inexpensive and can be easily processed into good optical quality thin films, which can be readily integrated into a large variety of devices [1-4]. The active chromophores orientation inside polymeric films is generally obtained by applying an external static field [5] or by all optical poling techniques [6].

The NLO effects of some push-pull organic molecules have been extensively investigated in order to incorporate them into materials devoted to technological devices, for example second-order optical non-linearities are useful in frequency doubling and optical switching with a fast response time [7-10]. The conjugated organic molecules, such as azo-dye molecules, have relatively large optical non-linear susceptibilities due to the de-localization of π-electronic clouds between groups of acceptors and donors of electrons.

---

* Contact author. Fax: +52(55)56225011 Phone: +52(55)56225103 E-mail: alfredofranco@fisica.unam.mx

Macroscopic optical second-order nonlinearities require a non-centrosymmetric orientation of the NLO molecules in bulk or in film materials. However, it is difficult to arrange these molecules in a macroscopically large crystalline form. An alternative approach in inducing macroscopic nonlinearities is the orientation of the electric dipole moments of the NLO guest molecules in a host polymer, using the Corona poling technique [5,11,12]. In many works on guest-host systems for second-order optical materials, all organic materials have been widely used because the molecular design and their modification are not extremely difficult [13,14]. From a practical point of view, the big issues to solve in all organic materials are their thermal instability and their low chemical resistance, which make them scarcely suitable for direct use in photonic devices. However, the possibility to increase the merits of this kind of materials by means of their nanostructuration has not been yet widely studied, but there are evidences that the presence of nanostructures in the materials can improve their nonlinear optical performances [15,16].

In this paper we present the kinetics of Second Harmonic Generation (SHG) for bottom-up nanostructured Polymethylmethacrylate (PMMA) films containing 4-(4-Nitrophenylazo)aniline (also named Disperse Orange 3 or simply DO3) chromophores. The DO3 azo dye was chosen as a guest of nonlinear optical films because its simple structure and their large linear and nonlinear polarizabilities, due to the existence of electron-donor and electron-acceptor groups attached to its extended double conjugated bonds system [17] (see Scheme 1). The Second Harmonic Generation of the polymeric film was studied as function of the Corona poling time at three different temperatures: 60ºC, 80ºC and 100ºC. In this work we name nanostructured material to that one whose matrix has a periodic long-range order with a $d$-spacing within the nanoscale range.

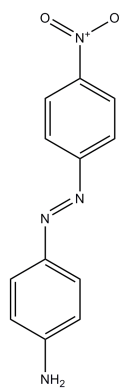

Scheme 1. Molecular structure of the azo dye 4-(4-Nitrophenylazo)aniline (Disperse Orange 3, DO3).

The chromophores orientation dynamics has been theoretically studied by J. W. Wu [18], D. J. Binks et. al. [19] and A. Franco et. al. [20,21]. We used the model developed in reference [21] for the analysis of our experimental results.

The knowledge of the temporal behavior of the SHG signal in these kinds of materials give information about the local interactions between the guest chromophores and their host film, which is important for the optimal design of nanophotonic devices.

## 2. EXPERIMENTAL

### 2.1 Materials synthesis

All the reactants were Aldrich grade and they were used as purchased. DO3 was the nonlinear optical chromophore in all the samples. All the samples were guest-host, and the DO3 concentration was the same for all the PMMA samples.

Two kinds of PMMA films were prepared: amorphous and nanostructured. Amorphous samples were prepared as follows: PMMA (F.W. = 120,000 g·mol$^{-1}$) was dissolved in Tetrahydrofuran (THF) and stirred magnetically for 15 minutes at room temperature, then DO3 was added and the mixture was stirred for 15 minutes. This final solution was filtered with a 0.45 µm pore size syringe filter.

80% of the total weight was liquid (THF), 20% was solid (PMMA+DO3). 95% of the solid weight was PMMA and 5% was DO3.

The films were deposited by dip-coating on microscope glass slides at a constant withdrawal speed of 20 cm/min. The films were annealed in air at 70ºC during 3 hours.

Nanostructured PMMA samples were prepared with a similar procedure, but 1.5 wt % of Sodium Dodecyl Sulfate (SDS), an anionic surfactant, was added just prior to filter. The final solution was stirred 5 minutes and then the films were deposited as described above.

**2.2 Materials characterization**

Long-range order of the polymeric matrix was identified by X-Ray Diffraction (XRD) measurements. The XRD patterns were recorded on a Bruker AXS D8 Advance diffractometer using Ni-filtered CuK$\alpha$ radiation. A step-scanning mode with a step of 0.02° in the range from 1.5º to 10°, in 2$\theta$, and an integration time of 2 seconds were used.

The SHG measurements were carried out in-situ using the experimental set-up shown in Figure 1. This set-up consists of a pulsed Nd:YAG (Nanolase NP-10620-100, wavelength: 1064 nm, frequency: 5 kHz, energy: 5µJ/pulse) as the source of the fundamental beam of light; two lenses, one of them focuses the fundamental laser beam on the sample, the second one collects the second harmonic light generated by the sample and sends it to a photomultiplier (Hamamatsu H5784) through a color filter which blocks the fundamental beam of light. The second lens was placed at its focal length distance from the sample. The photomultiplier was connected to an oscilloscope (Tektronix TDS 3052B) and the data were saved automatically in a computer each 0.5 seconds. The Corona field was produced by a high voltage (5 kV) between a silver needle and a copper plate separated each other by a distance equal to 1.2 cm. The films were hold on the copper plate electrode, which had attached a resistance that worked as a heater. The silver needle was disposed perpendicular to the copper plate.

The thickness of the films was measured using a scanning electron microscopy (SEM) by means of a JEOL JSM-5600-LV electron microscope.

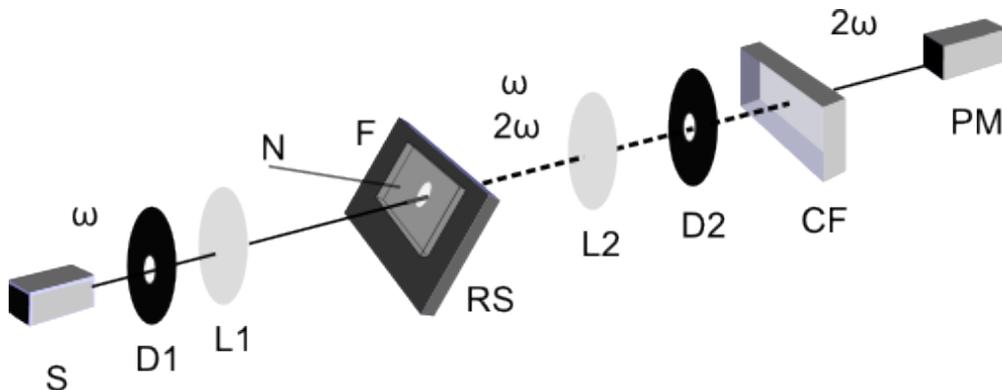

Figure 1. Schematic diagram of the SHG experimental setup. S: Nd:YAG laser (1064 nm) as source of the fundamental beam of light. D1, D2: Diaphragms. L1, L2: Convergent lenses. RS: Rotational stage with heater for Corona poling technique. F: Film. N: Silver needle for Corona poling technique. CF: Color filter. PM: Photomultiplier connected to an oscilloscope.

## 3. RESULTS AND DISCUSSION

The thickness measurements were carried out on several places of the samples. The thickness of the samples was very uniform, it was equal to 1.65 ± 0.08 µm (Figure 2), after 25 measurements.

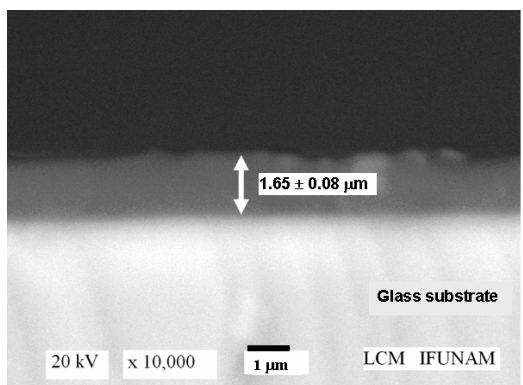

Figure 2. Image of a transversal section of a PMMA:DO3 film obtained by SEM for its thickness measurement.

## 3.1 UV-visible spectroscopy

Figure 3 shows the normalized optical absorption spectra of amorphous and nanostructured PMMA:DO3 films. The spectra were taken at room temperature in the range of 350-650 nm. All the main absorption bands of the spectra were placed at the same wavelength: 440 nm. But the spectra widths were not the same; in the nanostructured films the bands were broader.

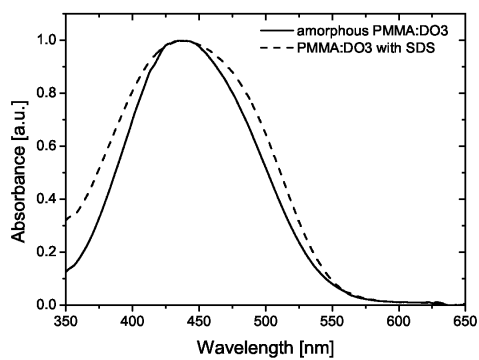

Figure 3. Normalized UV-visible optical absorption spectra of the PMMA:DO3 films.

## 3.2 XRD patterns

The XRD patterns obtained for the amorphous and nanostructured films appear in the Figure 4. The spectrum of the amorphous film does not show any peak. On the other hand, the XRD pattern of the films templated with SDS shows the peak (100) of a lamellar long-range order structure, with a *d*-spacing equal to 3.7 nm.

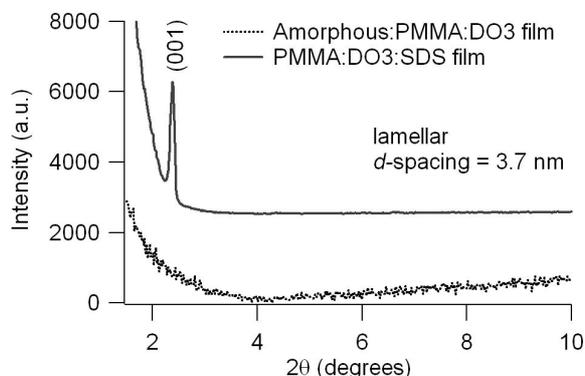

Figure 4. Small angle XRD patterns of the samples templated with SDS showing a lamellar nanostructure.

## 3.3 Order parameter

The comparison between the absorbance $A$ of the films, before and after a Corona poling process, quantifies the degree of orientation of the chromophores embedded in the films. This comparison, expressed as the percentage of the absorbance losses, is often called "order parameter" $A_2$, and it is determined by means of the next equation:

$$A_2 = 1 - \frac{A_\perp(t)}{A(t=0)}, \qquad (1)$$

where $t$ is the Corona poling time and the $\perp$ subscript indicates that the direction of the light used in the optical measurements is perpendicular to the face of the films, the same direction than the Corona field. In our case, the absorbance value was taken at the maximum of the main band, i.e. at 440 nm.

The $A_2$ parameter is often related to the percentage of chromophores oriented along the Corona poling field. It reaches a maximum value after enough large poling times; the value depends on the ability of the chromophores to get oriented inside the sample. From the $A_2$ maximum value of a film, its local electric field inside $E$ can be determined by means of the Rigid Oriented Gas Model, without considering any chromophore-chromophore interaction corrections [20]. Actually, larger $A_2$ values correspond to larger $E$ fields in the films.

The $A_2$ values obtained for the two kinds of samples (amorphous and nanostructured) at each temperature (60ºC, 80ºC and 100ºC) are shown in Figure 5 as $A_2$ vs. Corona poling time plots.

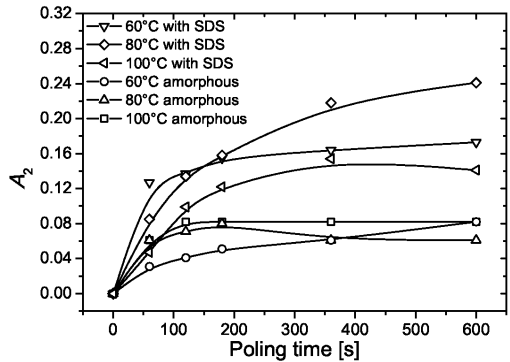

Figure 5. Plots of the $A_2$ order parameter as function of the Corona poling time. The continuous lines were drawn as visual aid.

## 3.4 Second Harmonic Generation

The SHG signal of the films grows as the poling time increases; the SHG signal of the films reaches a plateau at enough large poling times. This plateau occurs when the chromophores have reached their largest possible non-centrosymmetric arrangement (if the DO3 chromophores were arranged in a centrosymmetric way then the SHG signal would be equal to zero). The intensity of the SHG signal as function of the poling time is shown in Figure 6.

The experimental data were fitted using a previously reported model for the orientation of the chromophores [20]. The model uses as fitting parameters (1) a damping constant of the material $\gamma$, directly related to the chromophore-matrix interactions, and (2) a SHG intensity signal constant $C$ proportional to $(N\beta_{333}I^\omega)^2$, where $N$ is the number of non-linear optical active chromophores (DO3 in our case), $\beta_{333}$ is the second order hyperpolarizability of the chromophores and $I^\omega$ is the intensity of the fundamental beam of light. Thus, larger $\gamma$ values imply lower chromophores mobility, and larger SHG intensity constant values imply larger number of molecules contributing to the non-centrosymmetry of the material.

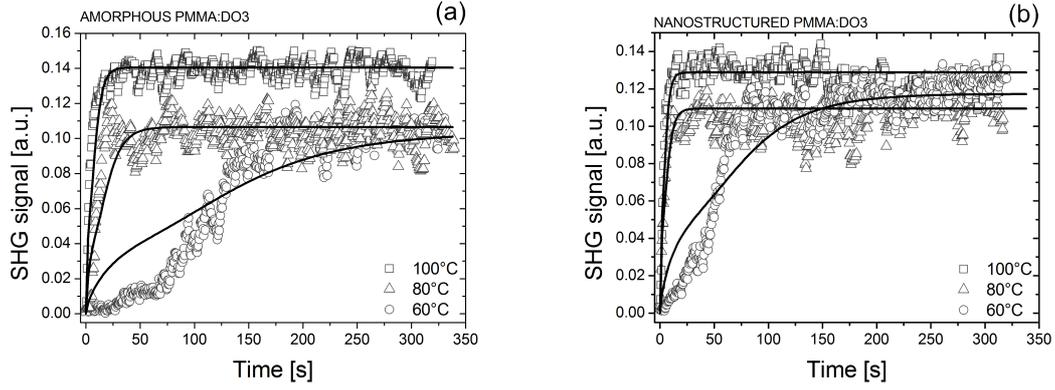

Figure 6. SHG signal as function of the Corona poling time for (a) amorphous, and (b) lamellar PMMA:DO3 films at three different temperatures, with their respective theoretical fits (black continuous lines).

Basically, the kinetics of each chromophore is described by a harmonic oscillator equation

$$\ddot{\theta} + 2\gamma \dot{\theta} + \omega^2 \theta = 0, \qquad (2)$$

where $\gamma$ is, in fact, the over-damping constant, $\omega$ is the natural frequency of chromophores with dipolar moment $\mu_3$ (2.47x10$^{-29}$ *C m* for DO3 [22]) and inertia moment $I_{33}$ (7.1x10$^{-44}$ *kg m$^2$* for DO3), under the effect of a local electric field *E*; the natural frequency is given by:

$$\omega = \sqrt{\frac{\mu_3 E}{I_{33}}}. \qquad (3)$$

Table 1 contains the parameters used to fit the experimental data of the Figure 6. The general form of the equation used for the fittings SHG signals is

$$I^{2\omega} = Cf(\theta(\gamma,\omega;t)), \qquad (4)$$

where $I^{2\omega}$ is the intensity of the SHG signal and *f* is a statistical function of all the possible angles of orientation $\theta$ of the chromophores, which depends of the $\gamma$ and $\omega$ parameters as well as the poling time *t* [20,21].

Table 1. Parameters used for the best fitting to the experimental results obtained for each studied material.

| SAMPLE | $\gamma$ (x10$^{25}$ s$^{-1}$) | SHG intensity constant *C* (a.u.) |
|---|---|---|
| Amorphous (60ºC) | 0.539 | 1.158 |
| Amorphous (80ºC) | 0.070 | 1.549 |
| Amorphous (100ºC) | 0.039 | 1.571 |
| With SDS (60ºC) | 0.502 | 0.697 |
| With SDS (80ºC) | 0.067 | 0.511 |
| With SDS (100ºC) | 0.033 | 0.891 |

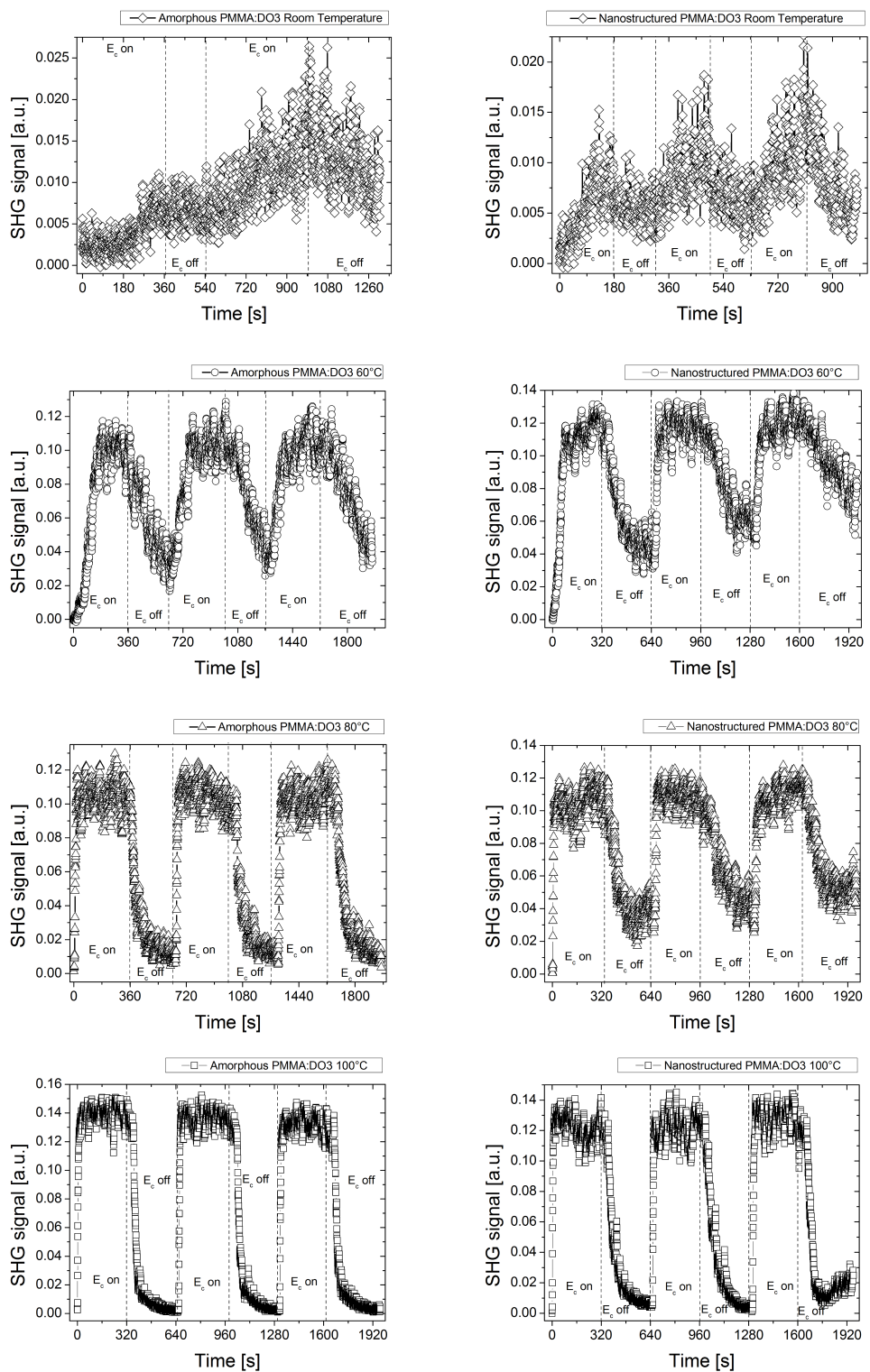

Figure 7. Cyclic plots of SHG signal intensity as function of the Corona poling time (E$_c$ on) and of the relaxation time (E$_c$ off) at room temperature, 60ºC, 80ºC and 100ºC.

In Figure 7 some cyclic measurements of the SHG are shown, these measurements were carried out to show the reproducibility of the measurements and in order to check the differences in the relaxation behavior between both kinds of samples.

In order to estimate the SHG signal decay rate when the Corona poling field is turned off, it is reported in Table 2 the time that each sample takes to relax its own SHG signal to the half of its initial intensity at each one of the temperatures.

Table 2. Time ($t_D$) at which the SHG signal of each sample reaches the half of its highest intensity, after turning off the Corona poling field, at four different temperatures.

| SAMPLE | $t_D$(s) |
|---|---|
| Amorphous (Room Temperature) | 298±25 |
| Amorphous (60ºC) | 145±26 |
| Amorphous (80ºC) | 56±27 |
| Amorphous (100ºC) | 62±6 |
| With SDS (Room Temperature) | 45±25 |
| With SDS (60ºC) | 185±46 |
| With SDS (80ºC) | 101±29 |
| With SDS (100ºC) | 79±9 |

It is important to mention that the $A_2$ values of Figure 5 are not altered by a possible degradation of the chromophores, because the SHG signals always reached their same maximum value after several identical poling cycles. Thus, in Figure 5, the maximum $A_2$ value for each sample was considered for determining the local electric field $E$ value and then for fitting the experimental results of Figure 6.

Tables 1 and 2 express the mobility of the chromophores during the growth of the SHG signal and during the decay of the SHG signal showed in Figure 7 as function of the temperature. Both, the growth and the decay of the SHG signal are faster as temperature increases. At each temperature, the growth of the SHG signal is faster in the nanostructured films, but the decay of the SHG signal is faster in the amorphous ones.

The amorphous samples have slightly larger $\gamma$ values than the nanostructured samples at the three different temperatures, it means that the chromophore non-centrosymmetric orientation is faster in the nanostructured films. But when the Corona poling field is turned off ($E_c$ off, Figure 7), the SHG signal of the nanostructured films decays slower than the amorphous ones. It indicates that the rise and the decay of the SHG signal give different kinds of information about the orientational kinetics of the chromophores.

As the poling temperature increases both, the relaxation decay rate (when the Corona field is off) and the over-damping parameter value $\gamma$, tend to be the same for both kinds of films. Actually, the long-range ordered structure detected by the XRD measurements in the samples templated with SDS disappears after a poling process at 100ºC.

The reverse situation occurs at room temperature, the amorphous films exhibit the largest SHG signal, but the nanostructured films show a faster relaxation when the Corona poling field is turned off.

It was not possible to make a good fitting to the SHG signal growth at room temperature, because the signal was too small and the growth behavior was very different to that one predicted by the theoretical models. Actually, the fittings showed in Figure 6 for the results obtained at 60ºC do not follow perfectly the experimental results, but at 80ºC and at 100ºC the theory fits quite well the experimental data. It is due to the fact that the over-damping parameter $\gamma$ of the theoretical model is always constant in time and temperature-independent. This fact is not necessarily true at close or lower temperatures than the glass transition temperature ($T_g$) of the films. Anyway, the theoretical fitting of the experimental data obtained at 60ºC was carried out only for reference purposes.

As we stated before, at 100ºC both kinds of samples are practically amorphous, but a couple of differences between the samples remain at this temperature: the SHG intensity constant is considerably larger in the samples without SDS, but the $A_2$ value is larger for the samples with SDS. A similar situation happens in the samples with SDS: the temperature at which $A_2$ takes its lowest value, is the temperature at which the SHG intensity constant takes its largest value, and viceversa (the temperature at which $A_2$ takes its largest value, is the temperature at which the SHG intensity constant takes its lowest value). It means that the SDS surfactant favors the chromophores orientation along the Corona poling field but in some way the SDS avoids that a large $A_2$ value be reflected as a large SHG signal.

At this point it is useful to remark that the SHG intensity constant $C$ do not reflects straightforwardly the experimental SHG signal intensity, because the calculation of the SHG intensity constant $C$ requires not only the experimental SHG signal intensity, but the maximum $A_2$ order parameter too ($A_2$ determines the local electric field $E$).

It would be expected that, as the temperature increases, the chromophores mobility increases too, and in consequence the non-centrosymmetric alignment of the chromophores can be done faster, just as the $\gamma$ values in Table I shows. But if the chromophores have an enough large free volume space, then their angular dispersion should be larger too. It means that neither the order parameter neither the SHG intensity increase monotonically as the temperature increases because the changes in temperature also affects to the backbone of the host film. It is remarkable the fact that at 80ºC the films with SDS reach their maximum $A_2$ value and the amorphous films reach their minimum $A_2$ value. Definitely the surfactants incorporation to the samples changes the temperature dependence of the host material behavior (in our case the behavior of the PMMA).

With respect to the SHG intensity constant, its value grows as the temperature increases for the amorphous samples, but the value has a minimum at 80ºC for the samples with SDS.

From the SHG cyclic measurements it is clear that, when the Corona poling field is turned off, the SHG signal decrease is slower in the nanostructured films.

## 4. CONCLUSIONS

High optical quality amorphous and nanostructured PMMA films doped with the chromophore DO3 were synthesized. A lamellar long-range order nanostructure was obtained in the PMMA films templated with SDS. Second harmonic generation experimental results as function of the Corona poling time in PMMA:DO3 guest-host films show some differences between amorphous and nanostructured samples in relation to their chromophores mobility and to their chromophores non-centrosymmetric alignment. The largest maximum SHG signal was measured for the amorphous PMMA films, and the lowest one was measured for the nanostructured PMMA films. The shortest rise time for the SHG signal was observed in the nanostructured films, and the largest one was obtained in the amorphous films. The fastest decay time for the SHG signal, when the Corona poling field is off, was observed in the amorphous films; and the largest one was obtained in the nanostructured films.

## ACKNOWLEDGEMENTS

CONACYT 79781, NSF-CONACYT, PUNTA, PAPIIT 107510 and UNAM-UNIPD agreement supported this work. The authors also thank to M. in Sci. Manuel Aguilar-Franco (XRD measurements), Jaqueline Cañetas Ortega (SEM measurements) and Diego Quiterio (preparation of the samples for SEM) for technical assistance.